\def\beg{\begin{equation}}
\def\eeq{\end{equation}}

\documentstyle[12pt]{article}
\begin{document}

\pagestyle{headings}
\begin{center}
{\Large{\bf NMR in the quantum Hall effect}}
\vskip1.35cm
{\bf Keshav N. Shrivastava}
\vskip0.35cm
{\it School of Physics, University of Hyderabad,\\
Hyderabad   500 046, India.}
\end{center}
\vskip0.5cm
{\bf Abstract.}\\

We report that the rate of field sweep is important for
the observation of the zeroes in the transverse resistivity
in the quantum Hall effect. The resistivity also shows 
resonances at the nuclear magnetic resonance (NMR) frequencies when the 
r.f. coil is placed along the x direction. The relationship
of the eigen values with the resistivity is pointed out which shows
that even if the eigen values are largely corresponding to
that of the single-particle, the plateau widths, shifts and 
shape require many-body interactions.

\vskip0.9cm
\newpage
\baselineskip22pt
\noindent Recently, Kronm\"uller et al$[1]$ have found that the
longitudinal resistivity shows maxima in samples of
GaAs/Al$_x$Ga$_{1-x}$As which have small well thickness and when
the field sweep rate is small, 0.002 T/min, instead of zeros at
the same values of the magnetic field at which the transverse
resistivity, $\rho_{xy}$ shows plateaus. Usually, at a large
sweep rate in samples of large thickness, the longitudinal
resistivity, $\rho_{xx}$, shows zeros at certain values of the
field, $B_z$, at which $\rho_{xy}$ shows plateaus. The optically
pumped$[2]$ NMR studies have been performed at the Landau level
filling factor of $\nu=1/2$ which show that at the points where
$\rho_{xx}$ shows zeros and $\rho_{xy}$ plateaus, the maxima in
$\rho_{xx}$ occur instead of zeros and the existence of field is
proved by performing the NMR of $^{71}$Ga. According to Maxwell
equations, the zero in the resistivity is consistent with the
zero in the magnetic field. However, in the quantum Hall effect
$\rho_{xx}=0$ points have large finite fields, $B_z$ which
requires that ordinary Maxwell equations may be modified to
accomodate the experimentally observed fields when
$\rho_{xx}=0$. The consistency in the Maxwell equations may be
obtained by a suitable shift in the field. The entire field is
subject to the flux quantization but it is not necessary to
split it into two terms, one flux quantized and the other not
quantized. It is also not necessary to attach fluxex 
even number or odd, to the electrons.

In the present letter, we find that both the electric as well as
the magnetic fields require modification and not only the vector
potential but also the scalar potential has to be corrected to
understand the NMR at such points where $\rho_{xy}$ has plateaus
and $\rho_{xx}$ has zeros. Since the NMR shows that the magnetic
field is not zero, it is clear that $\rho_{xx}$ should also not
be zero. Therefore, we report modifications in both the electric
as well as the magnetic fields and hence in the scalar as well
as the vector potential of the electromagnetic fields.

In quantum Hall effect, the magnetic field is applied along the
$z$ direction, the electric field along the $x$ direction and
the Hall voltage is measured along the $y$ direction, in a
layered semiconductor with layers in the $x-z$ plane. It is
found that the longitudinal resistivity, $\rho_{xx}$, as a
function of magnetic field, $B_z$, becomes zero at several
values of $B_z$ and oscillates between zeros and maxima. The
transverse resistivity in the $xy$ plane, $\rho_{xy}$, shows
plateaus at the same values of field, $B_z$, at which
$\rho_{xx}$ shows zero values. In the samples of
GaAs/Al$_x$Ga$_{1-x}$As with reduced well thickness, 15 nm, as
compared with the conventional ones, the transverse resistivity
shows maxima, at the same values of the fields where zeros occur
in the large well samples. At integer filling factors the
resistance, $\rho_{xx}$ goes to zero and at filling factor
$\nu=2/3$ there is a minimum but when the sweep rate of the
magnetic field is reduced to 0.002 T/min a huge maximum (HLR) is
found instead of a minimum or zero. Kronm\"uller et al$[1]$ have
performed the nuclear magnetic resonance measurement of
$^{75}$As at the same field as that at which $\rho_{xx}$ is a
maximum. Thus it is proved that there is a field at which
$\rho_{xx}=0$. We make an effort to understand the NMR at a
field at which there is a huge longitudinal resistance (HLR).
This is the same point at which $\rho_{xx}$ shows zero in thick
well samples and $\rho_{xy}$ shows plateaus. In superconductors,
the zero resistivity is consistent with zero field but in the
present case there is always a field along $z$ direction.
Therefore, we develop a theory which can give a zero resistivity
at a finite magnetic field without superconductivity. Such a theory
 does not exist in the literature.

We assume that magnetic induction is independent of time so that
$dB/dt=0$. According to a Maxwell equation,
\beg
\nabla \times E = -{\partial B\over\partial t}
\eeq
so that $\nabla\times E=0$, which gives $E=0$. Since the current
$j=(1/\rho)E$, the value $E=0$ is consistent with $\rho=0$. This
is true in superconductivity where all components of $B$ are
zero. In the quantum Hall effect, (QHE) when $\rho_{xx}=0$, there
is a large magentic field along the $z$ direction. Therefore
$B=0$ and $\rho_{xx}=0$ type theory is not applicable. In order
to understand the QHE, we suggest that both $E$ and $B$ are
shifted. The quasiparticles of charge $\nu e$ with a finite
charge density, produce a field due to Maxwell equations
\begin{eqnarray}
\nabla\cdot E_o&=&4\pi \rho_o\nonumber\\
\nabla\times B_o-c^{-1}(\partial E_o/\partial t)&=&(4\pi/c)j
\end{eqnarray}
so that we can assume that there are fields $E_o$ and $B_o$
which shift the electric field and the magnetic induction. Now,
$B_z$ is replaced by $B_z-B_o$ and
\beg
{d\over dt}(B_z-B_o) = 0
\eeq
so that there is no field inside the semiconductor. Substituting
this result into the Maxwell equation and replacing $E_x$ by
$E_x-E_o$, we find,
\beg
\nabla \times (E_x - E_o) = 0
\eeq
so that $E_x-E_o=0$. The Ohm's law is
\beg
\rho_{xx}j=E_x - E_o
\eeq
which is consistant with $\rho_{xx}=0$ for $E_x-E_o=0$. This
result gives the usual QHE. The magnetic induction inside the
semiconductor is
\beg
B_z - B_o + 4\pi M = 0
\eeq
because $\rho_{xx}=0$. Here $M$ is the magnetization of the
sample. The above result shows that the points where
$\rho_{xx}=0$ have diamagnetic magnetization, 
\beg
{M\over B_z-B_o} = -{1\over 4\pi}\,\,.
\eeq
The $\rho_{xx}=0$ is thus consistant with a shift $B_o$ in the
field. Since there is a sweep rate dependence, the above
processes occur in a time $\tau$ such that the field shift is
given by
\beg
B_o = \tau {d\over dt} B_o = \tau\alpha_o
\eeq
where $\alpha_o$ is the sweep rate. The effective field is thus
given by 
\beg
B_{eff} = B_z - \tau {d\over dt} B_o = B_z -\tau\alpha_o\,\,\,.
\eeq
For small values of $\alpha_o$, the condition
\beg
f(\alpha_o) = B_z -\tau\alpha_o + 4\pi M = 0
\eeq
is not satisfied so that $\rho_{xx}=0$ points do not occur. Thus
the maximum value of $\alpha_o$ for which $f(\alpha_o)=0$ is 
\beg
\alpha_{o,min} = (B_z + 4\pi M)/\tau\,\,\,.
\eeq
If $\alpha_o$ is less than $\alpha_{o,min}$, the value of
$f(\alpha_o)$ is positive and $\rho_{xx}=0$ does not occur. Large
values of $\alpha_o$ give negative $f(\alpha_o)$ while small
values of $\alpha_o$ give positive $f(\alpha_o)$. Thus there is
a particular value of $\alpha_o$ for which $\rho_{xx}=0$ points
occur. For small values of $\alpha_o$ finite value of
$\rho_{xx}$ occurs. Thus the quantum Hall effect requires an
effective charge $\nu e$ of the quasiparticles which shifts both
$E$ and $B$ and the sweep rate of the field is important for
the observation of zeros in the $\rho_{xx}$ at certain finite
values of the magnetic field along the $z$ direction. Since
$\vec{B}=\vec{\nabla}\times\vec{A}$ a shift in $\vec{B}$ amounts
to a shift in the vector potential $\vec{A}$, but there is a
time dependent term in the above so that $\partial A/\partial t$
is not zero. Since the scalar potential $\phi$ depends on $E$ as
well as on $\partial A/\partial t$, by means of the relation
$E+(1/c)(\partial A/\partial t)=-\nabla\phi$, it is clear that
$\phi$ also requires a correction due to shift $E_o$ in $E$ and
the finite sweep rate dependence in $B$ and hence in $\partial
A/\partial t$. 

According to the Chern-Simons
transformation$[3,4]$ in three dimensions, one can add a term
$\epsilon_{ijk}A_iF_{jk}$ to the vector potential $\vec{A}$ in
the expression for the linear momentum $p-{e\over c}\vec{A}$
without disturbing the gauge and Lorentz invariance. Our
calculation clearly shows that additional shift in the magnetic
field is consistent with the Chern-Simons (CS) transformation
and hence the vector potential is shifted. However, it is also
clear that the scalar potential should also be corrected. Thus
not only the vector potential (CS) but also the scalar potential
is shifted.

For $\alpha_o < \alpha_{o,min}$ the $\rho_{xx}=0$ point does not
occur but the $\rho_{xy}$ has plateaus. The field $B_z$ at this
point may be called $B_{eff}$ so that NMR may be used to measure
this field. The radio-frequency coil is fixed such that the r.f.
oscillating field is along the $x$ direction with $B_z$ along
the $z$ direction. The NMR transition occurs when the resonance
frequency matches with the effective field,
\beg
\omega = \gamma B_{eff}
\eeq
where $\gamma=g_N\mu_N/\hbar$ is the nuclear gyromagnetic ratio.
Thus the NMR is shifted according to eq.(9). The quadrupole
interaction is given by the hamiltonian,
\beg
{\cal H} = g_N\mu_N B_{eff}.I_z +
Q^\prime[I^2_z-{1\over3}I(I+1)] + Q^{\prime\prime}(I^2_x-I^2_y)
\eeq
where $g_N$ is the nuclear $g$ factor, $\mu_N$ is the nuclear
magneton, $I$ is the nuclear spin and $Q^\prime$ and
$Q^{\prime\prime}$ are the nuclear quadrupole interaction
constants. Usually, the transitions $-3/2 \to -1/2, -1/2\to+1/2$
and $1/2\to 3/2$ are superimposed on each other so that only one
spectral line occurs. However, in the present case $-3/2\to1/2$
and $-1/2\to3/2$ transitions are also quite strong. The
separation between these two transitions is
\begin{eqnarray}
E_{-3/2\to1/2} - E_{-1/2\to3/2} &=& 2[(Q^\prime + g_N\mu_N
B_{eff})^2 + 3Q^{\prime\prime2}]^{1/2}\nonumber\\
&& + [(Q^\prime-g_N\mu_N B_{eff})^2 + 3Q^{\prime\prime2}]^{1/2}\,\,\,.
\end{eqnarray}
In addition $-3/2\to +3/2$ is also possible so that there are
four lines in the NMR spectrum of $^{75}$As with $I=3/2$ as
clearly seen in the spectra of $\rho_{xx}$ as a function of r.f.
frequency. In the NMR experiments, it is sufficient to have a
large field along the $z$ direction and a small r.f. oscillating
field along the $x$ direction. However, in the present case,
there is a large electric field along the $x$ direction.
Therefore, the theory of usual NMR requires correction for the
electric field. The wave functions are mixed
$|a^\prime>=a|3/2>+b|1/2>+c|-1/2>+d|-3/2>$ due to interactions
of the form $\sum_i\sum_{j\le
k}({1\over2})R_{ijk}E_i(I_jI_k+I_kI_j)$ where $E_i$ are the
components along the $x,y$ and $z$ axis which correct the
nuclear quadrupole interaction as given in an analogous
problem$[5,6]$. The relaxation at very low temperatures such as
0.3 K is caused by the absorption and emission of photons which
gives rise to very long relaxation times, of the order of
minutes. The details of the calculation of relaxation times were
given long time ago. The relative change in resistivity is given
by,
$$
{\delta R_{xx}\over R_{xx}} = \left({L_o\omega\over R_o}\right)
4\pi\chi^{\prime\prime} = 4\pi \chi^{\prime\prime} Q
$$
where the $\chi^{\prime\prime}$ is the imaginary part of the
susceptibility,
$\chi=\chi^{\prime}(\omega)-i\chi^{\prime\prime}(\omega)$, $Q$
is the quality factor of the coil and the inductance is
$L_o(1+4\pi\chi_o)$. The coil of inductance $L_o$ is filled with
the material of susceptibility, $\chi_o$. The imaginary part of
the susceptibility is given by,
$$
\chi^{\prime\prime} = {\chi_o\over2} \omega_o T_2
{1\over1+(\omega-\omega_o)^2T^2_2} 
$$
where $T_2$ determines the life time which in this case is
caused$[7]$ by the radiative process. At $\omega=\omega_o$ there
is a resonance detected by the resistivity, $\rho_{xx}$. The
hyperfine splitting has not been seen by Kronm\"uller et al$[1]$.
However, in general the isotropic value of the hyperfine
constant, $A$, in the hyperfine interaction A.I.S is determined by
$$
A_s = {8\pi\over3} gg_N\mu_B\mu_N |\psi(o)|^2\,\,\,.
$$
If $e$ is changed to $\nu e$, the change may be introduced as
in previous studies $[8,9]$.  Hence it will be of interest to
observe hyperfine interactions with fractional charge.
Kronm\"uller et al have thus opened the doors to a whole variety
of new NMR measurements.

It turns out that the values of the fractions given in Fig.18 
of St\"ormer's Nobel lecture [10] are the same as those given
by Shrivastava[11] in 1986. The equality of masses of some
of the quasiparticles is also well explained [8] by the
mechanism of Shrivastava [11]. The high Landau levels are also
understood by this mechanism [12,13]. Considerable amount of
the experimental data agrees very well with Shrivastava's
theory [14]. It is found that there is a correction to the
value of the Bohr magneton[15].
From this study it is clear that as far as determining the
centres of the plateus in the $\rho_{xy}$ and the zeroes in
$\rho_{xx}$ is concerned, there is little role of the 
many-body theory. However, electrons interact with radiation
and with phonons, so that many-body theory enters for the
determination of the plateau length and shape of the minima curves.
 
In conclusion, we find that both the electric as well as the
magnetic fields are subject to a shift in the quantum Hall
effect so that a new mode is predicted. The NMR can be
 detected by measuring the resonances in the resistivity, $\rho_{xx}$.

\newpage
\parindent0mm
\begin{enumerate}
\item  S. Kronm\"uller,  W. Dietsche, K. v. Klitzing,  G. Denninger,
	W. Wegscheider W and  M. Bichler,Phys. Rev. Lett. {\bf
	82}, 4070 (1999).
\item A. E. Dementyev,  N. N. Kuzma,  P. Khandelwal,  S. E. Barrett,
      L. N. Pfeiffer and K. W. West,  Phys. Rev. Lett. {\bf 83},
	5074 (1999).
 \item S. S. Chern and  J. Simons,  Proc. Natl. Acad. Sci. (U.S.A.)
	{\bf68}, 791 (1971).
\item A. Stern A, B. I. Halperin, F. v. Oppen F v and S. H. Simon S H,
      Phys. Rev. B{\bf59}, 12547 (1999).
\item A. Kiel A and  W. B. Mims, Phys. Rev. B{\bf5}, 803 (1972)
\item A. B. Roitsin, Usp. Fiz. Nauk {\bf105}, 677 (1971); [Sov.
	Phys. Uspekhi {\bf14}, 766 (1972)].
\item K. N. Shrivastava, Phys. Stat. Solidi B{\bf115}, K 37 (1983).
\item K. N. Shrivastava, Mod. Phys. Lett. B{\bf13}, 1087 (1999).
\item K. N. Shrivastava, Superconductivity: Elementary Topics,
	World Scientific Pub. Co.: Singapore; River Edge, N. J., (2000).
\item H. L. St\"ormer, Rev. Mod. Phys. {\bf71}, 875 (1999).
\item K. N. Shrivastava, Phys. Lett. A {\bf113}, 435 (1986).
\item K. N. Shrivastava, Mod. Phys. Lett. {\bf14}, 1009 (2000).
\item K. N. Shrivastava, cond-mat/0103604.
\item K. N. Shrivastava, CERN SCAN-0103007.
\item K. N. Shrivastava, cond-mat/0104004. 
\end{enumerate}

\end{document}